\documentclass[a4paper,11pt]{article}
\pdfoutput=1 

\usepackage{jheppub}
\usepackage{epsfig}
\usepackage{amsmath}
\usepackage{graphicx}
\newcommand{\un}[1]{{\mathrm{\,#1}}}

\newcommand{\GeV}{\un{GeV}}

\newcommand{\lrf}[2]{ \left(\frac{#1}{#2}\right)}

\def\A0#1{\Pi_{\rm #1}(0)}
\def\AP0#1{\Pi'_{\rm #1}(0)}
\def\be{\begin{equation}}
\def\ee{\end{equation}}
\def\bea{\begin{array}}
\def\eea{\end{array}}
\def\beqa{\begin{eqnarray}}
\def\eeqa{\end{eqnarray}}
\def\beqas{\begin{eqnarray*}}
\def\eeqas{\end{eqnarray*}}

\def\bp{\begin{picture}}
\def\ep{\end{picture}}
\def\bc{\begin{center}}
\def\ec{\end{center}}
\def\bfig{\begin{figure}}
\def\efig{\end{figure}}

\def\bit{\begin{itemize}}
\def\eit{\end{itemize}}
\def\nn{\nonumber}
\def\f{\frac}

\def\[{\left[}
\def\]{\right]}
\def\({\left(}
\def\){\right)}

\def\..{\left.}
\def\.{\right.}
\def\tl{\tilde}

\def\la{\leftarrow}

\def\tm{\times}

\def\la{\lambda}

\def\al{\alpha}

\def\ep{\epsilon}

\def\ga{\gamma}
\def\pa{\partial}

\title{Explaining The Muon $g-2$ Anomaly and New CDF II W-Boson Mass in the Framework of (Extra)Ordinary Gauge Mediation}
\author[a]{Xiao Kang Du,}
\author[a]{Zhuang Li,}
\author[a]{Fei Wang,}
\author[a]{Ying Kai Zhang,}

\affiliation[a]{School of Physics, Zhengzhou University, Zhengzhou 450000, P. R. China}
\emailAdd{feiwang@zzu.edu.cn}
\abstract{The SUSY contributions $\Delta a_\mu$ to muon $g-2$ anomaly can not even reach $3\sigma$ in ordinary gauge mediated SUSY breaking (GMSB) scenarios because of the strong correlations between the colored sparticle masses and the uncolored EW sparticle masses. An interesting extension to GMSB is the (Extra)Ordinary Gauge Mediation (EOGM), which can relax the correlations between squarks and sleptons with non-universal choices for $N_{eff,3}$ and $N_{eff,2}$. We find that EOGM scenarios with $N_{eff,3}\ll N_{eff,2}$ can explain the muon $g-2$ anomaly within $3\sigma$ range, however can not explain the new W-boson mass by CDF II. We also propose to extend EOGM with additional adjoint $\Sigma_8$ and $\Sigma_3$ messengers at a high scale of order $1.0\times 10^{14}$ GeV, which can shift the gauge coupling unification
scale to the string scale. Such EOGM extension scenarios with adjoint messengers could spoil the unwanted gaugino mass ratios and give large SUSY contributions to $\Delta a_\mu$ for $N_{eff,3}\ll N_{eff,2}$, which can explain the muon $g-2$ anomaly up to $1\sigma$. Besides, because of the large messenger scale of order $1.0\times 10^{14}$ GeV, such scenarios will in general lead to large $|A_t|$ at the EW scale, which can accommodate the 125 GeV Higgs easily and possibly lead to smaller EWFT as well as BGFT. We discuss the possibility to explain the new CDF II W-boson mass in the GMSB-type framework. We find that SUSY contributions can marginally account for the new W-boson mass in the region with sleptons and wino both being light. }

\begin{document}
\maketitle
\newpage
\section{Introduction}
The predictions of the standard model (SM) has already been corroborated by
various contemporary collider experiments, including the discovery of Higgs boson by the Large Hadron Collider (LHC)~\cite{ATLAS:higgs,CMS:higgs}. Nevertheless, some experiments have found several subtle deviations from the SM predictions that remain to be accounted for, for example, the anomaly of muon anomalous magnetic moment $a_\mu\equiv (g-2)_\mu/2$ and the measurement of W boson mass. Combining the recent reported E989 muon $g-2$ measurement with the previous BNL result~\cite{g-2:BNL,g-2:PDG}, the update experimental value of $a_\mu$ is given by~\cite{g-2:FNAL}
\beqa
a^{\rm FNAL+BNL}_\mu = (11659206.2 \pm 4.1) \tm 10^{-10}~.
\eeqa
The world average deviation from the SM prediction~\cite{g-2:Th}
 \beqa
\Delta a^{\rm{FNAL+BNL}}_\mu =(25.1 \pm  5.9)  \tm 10^{-10}~,
 \eeqa
is pushed to a significance of $4.2\sigma$ with this new experimental value. Besides, the measurement of the W boson mass can provide a stringent test of the standard model. Using data corresponding to $8.8 fb^{-1}$ of integrated luminosity collected in proton-antiproton collisions at a 1.96 TeV  center-of-mass energy, the new value of W boson mass can be obtained to be
\beqa
M_W=80,433.5 \pm  6.4({\rm stat}) \pm 6.9 ({\rm syst})=80,433.5\pm 9.4 {\rm MeV}/c^2~,
\eeqa
by the CDF II detector at the Fermilab Tevatron collider~\cite{CDF:W}. This measurement is in significant tension with the standard model expectation which gives~\cite{SM:W}
\beqa
M_W=80,357\pm 4({\rm inputs})\pm 4 ({\rm theory}) {\rm MeV}/c^2~.
\eeqa
Such deviations, if persist and get confirmed by other experiments, will strongly indicate the existence of new physics beyond SM~\cite{Athron:muong-2,anomaly:W}, for example, the weak scale supersymmetry (SUSY).

  Weak scale SUSY is the leading candidate for new physics beyond the SM which can tame the quadratic divergence of the Higgs boson and realize successfully the gauge coupling unification by the contributions of new superpartners. The discovered 125 GeV Higgs scalar also lies miraculously in the small $'115-135'$ GeV window predicted by the low energy SUSY. The dark matter puzzle and the origin of baryon asymmetry in the universe can also be explained in the framework of SUSY. So, if low energy SUSY is indeed the new physics beyond the SM, it should account for the muon $g-2$ anomaly and the new measurement of W-boson mass.

However, low energy SUSY confronted many challenges from recent collider and DM experiments~\cite{GAMBIT}.  In order to be consistent with various experimental results and keep as many virtues of SUSY as possible, low energy SUSY should have an intricate structure, especially its soft SUSY breaking parameters, which are determined by the SUSY breaking mechanism. In attempting to construct SUSY models, it is necessary to include additional dynamics to  spontaneously break SUSY. The simplest possibility is to adopt an O'Raifeartaigh model and assume that some interactions can transmit the breaking of SUSY in the hidden sector to the ordinary matter fields.
 Depending on the way the visible sector $'feels'$ the SUSY breaking effects in the hidden sector, the SUSY breaking mechanisms can be classified into gravity mediation~\cite{SUGRA}, gauge mediation~\cite{GMSB}, anomaly mediation~\cite{AMSB} scenarios, etc. Gauge mediated SUSY breaking (GMSB) mechanism, which would not cause flavor and CP problems that bothers gravity mediation models, are calculable, predictive, and phenomenologically distinctive with minimal messenger sector. However, unless extended (for example by messenger-matter interactions~\cite{GMSB:MM}), minimal GMSB can hardly explain the 125 GeV Higgs with TeV scale soft SUSY breaking parameters because of the vanishing trilinear terms at the messenger scale.
 An interesting extension of minimal GMSB is the (Extra)Ordinary gauge mediation (EOGM) scenarios~\cite{EOGM}, in which the messenger sector can include all renormalizable, gauge invariant couplings between the messengers and any number of singlet fields. In fact, many examples in the literature of ordinary gauge mediation deformed by mass terms can fall into this category and their generic properties can be obtained therein.

Unlike the mass relations between squark and slepton in minimal gauge mediation (MGM),
the relevant mass relations can be modified randomly in EOGM. The effective number of messengers can be a continuous function of the couplings, taking values between $0$ and $N$ (the number of messengers in ${\bf 5}\oplus\bar{\bf 5}$ of $SU(5)$) inclusive. Such a modification to minimal gauge mediation predictions are welcome because the resulting spectrum can possibly accommodate the 125 GeV Higgs, the collider constraints. Amended mass relations for squarks and sleptons in EOGM may also lead to light sleptons and electroweakinos, which can give large contributions to the $\Delta a_\mu$ and sizeable corrections to W-boson mass via loops.

The gaugino mass ratios $M_1:M_2:M_3\approx 1:2:6$ will always be kept in GMSB and EOGM. Such gaugino ratios could be unfavorable to explain the previous mentioned anomaly, given the stringent LHC constraints on gluino mass. So, it is desirable to spoil such gaugino mass ratios to relax the lower mass bounds for electroweakinos. It had been shown that additional adjoint messengers $\Sigma_{3;b}({\bf 1,3,0})\oplus \Sigma_{8;b}({\bf 8,1,0})$ lie between $10^{12}$ GeV and $10^{14}$ GeV can still preserve gauge coupling unification~\cite{Mess:adjoint,HT:adjoint}. Such high scale new adjoint messengers can be adopted in GMSB to relax the low scale gaugino ratios and at the same time increase the low scale value of $|A_t|$, which are welcome to explain the discrepancies. We propose to include such adjoint messengers in EOGM to relax the unwanted gaugino ratios.

This paper is organized as follows. In Sec~\ref{sec-2}, general discussions on EOGM explanations of muon $g-2$ anomaly are given. In Sec~\ref{sec-3}, we survey the scenarios of EOGM extension with high scale adjoint messengers to see if the muon $g-2$ anomaly and CDF II W-boson mass can be explained in our scenarios. Sec~\ref{sec-4} contains our conclusions.

\section{\label{sec-2} The solution to $g_\mu-2$ and $\Delta m_W$ in ordinary EOGM model}

It is well known that the SUSY contributions to the muon $g-2$ are dominated by the chargino-sneutrino and the neutralino-smuon loops. At the leading order of $\tan\beta$ and $m_W/m_{SUSY}$, with $m_{SUSY}$ the SUSY-breaking masses,
 various loop contributions can be approximately give by~\cite{moroi}
 \begin{align}
 \Delta a_{\mu }(\tilde{W}, \tilde{H}, \tilde{\nu}_\mu)
 &= \frac{\alpha_2}{4\pi} \frac{m_\mu^2}{M_2 \mu} \tan\beta\cdot
f_C
 \left( \frac{M_2 ^2}{m_{\tilde{\nu }}^2}, \frac{\mu ^2}{m_{\tilde{\nu }}^2}  \right) ,
 \label{eq:WHsnu} \\
 \Delta a_{\mu }(\tilde{W}, \tilde{H},  \tilde{\mu}_L)
 &= - \frac{\alpha_2}{8\pi} \frac{m_\mu^2}{M_2 \mu} \tan\beta\cdot
 f_N
 \left( \frac{M_2 ^2}{m_{\tilde{\mu }_L}^2}, \frac{\mu ^2}{m_{\tilde{\mu }_L}^2} \right),
 \label{eq:WHmuL}  \\
 \Delta a_{\mu }(\tilde{B},\tilde{H},  \tilde{\mu }_L)
  &= \frac{\alpha_Y}{8\pi} \frac{m_\mu^2}{M_1 \mu} \tan\beta\cdot
 f_N
 \left( \frac{M_1 ^2}{m_{\tilde{\mu }_L}^2}, \frac{\mu ^2}{m_{\tilde{\mu }_L}^2} \right),
 \label{eq:BHmuL} \\
  \Delta a_{\mu }(\tilde{B}, \tilde{H},  \tilde{\mu }_R)
  &= - \frac{\alpha_Y}{4\pi} \frac{m_{\mu }^2}{M_1 \mu} \tan \beta \cdot
  f_N \left( \frac{M_1 ^2}{m_{\tilde{\mu }_R}^2}, \frac{\mu ^2}{m_{\tilde{\mu }_R}^2} \right), \label{eq:BHmuR} \\
 \Delta a_{\mu }(\tilde{\mu }_L, \tilde{\mu }_R,\tilde{B})
 &= \frac{\alpha_Y}{4\pi} \frac{m_{\mu }^2 M_1 \mu}{m_{\tilde{\mu }_L}^2 m_{\tilde{\mu }_R}^2}  \tan \beta\cdot
 f_N \left( \frac{m_{\tilde{\mu }_L}^2}{M_1^2}, \frac{m_{\tilde{\mu }_R}^2}{M_1^2}\right), \label{eq:BmuLR}
\end{align}
with $m_\mu$ being the muon mass and $\mu$ the Higgsino mass, $\alpha_Y=g_1^2/4\pi$ and $\alpha_2=g_2^2/4\pi$ for the $U(1)_Y$ and  $SU(2)_L$ gauge groups, respectively. The loop functions are defined as
\begin{align}
&f_C(x,y)= xy
\left[
\frac{5-3(x+y)+xy}{(x-1)^2(y-1)^2}
-\frac{2\log x}{(x-y)(x-1)^3}
+\frac{2\log y}{(x-y)(y-1)^3}
\right]\,,
\\
&f_N(x,y)= xy
\left[
\frac{-3+x+y+xy}{(x-1)^2(y-1)^2}
+\frac{2x\log x}{(x-y)(x-1)^3}
-\frac{2y\log y}{(x-y)(y-1)^3}
\right]\,,
\label{moroi3}
\end{align}
which satisfy $0\le f_{C,N}(x,y) \le 1$ and are monochromatically increasing for $x>0$ and $y>0$.
In the limit of degenerate masses, the loop functions satisfy $f_C(1,1)=1/2$ and $f_N(1,1)=1/6$. The SUSY contributions to the muon $g_\mu-2$ will be enhanced for small soft SUSY breaking masses and large $\tan\beta$.

  Generally speaking, large SUSY contributions to $\Delta a_\mu$ require light sleptons and light electroweakinos. However, soft SUSY spectrum of this type is difficult to realize when the collider and DM constraints etc are taken into account.
 In fact, weak scale SUSY is facing many challenges from the LHC experiments, especially the null search results of superpartners.  Recent analysis based on the Run-2 of 13 TeV LHC data of 139 fb$^{-1}$ integrated luminosity~\cite{Run2} constrain the gluino mass $m_{\tl{g}}$ to lie above 2.2 TeV~\cite{CMSSM:gluino} and the top squark mass $m_{\tl{t}_1}$ above 1.1 TeV~\cite{CMSSM:stop} in the context of simplified models. The discovered 125 GeV Higgs also impose stringent constraints on certain soft SUSY breaking parameters, such as the stop masses and the value of $A_t$. In fact, after imposing various LHC constraints, the popular CMSSM/mSUGRA with universal inputs at the GUT scale can not explain the muon $g-2$ anomaly~\cite{Wang:2021bcx}.  The region which can account for the recent $g-2$ anomaly is excluded by LHC direct searches for sparticles (and also be constrained stringently by Br($b \to s \ga$) bound). Besides, the region which can explain the 125 GeV Higgs has no overlap with the survived region that can account for the recent $g-2$ anomaly~\cite{Wang:2021bcx}. Some non-universal gaugino extension scenarios of CMSSM/mSUGRA, for example, the gluino SUGRA scenario, can elegantly explain the muon $g-2$ anomaly~\cite{Li:2021pnt}.

 In the predictive MGM models, the soft SUSY breaking parameters at the messenger scale are given by
\beqa
M_i=\f{\al_i}{4\pi}\f{F}{M}N_{\bf 5}~,~\quad {m}^2_{\tl{f}}=2\sum\limits_{i=1}^2C_{\tl{f}}^i \(\f{\al_i}{4\pi}\)^2\(\f{F}{M}\)^2 N_{\bf 5}~,
\eeqa
with $C_{\tl{f}}^i$ the corresponding quadratic Casimir invariants. The gaugino masses at the messenger scale satisfy
\beqa
\f{M_1}{g_1^2}=\f{M_2}{g_2^2}=\f{M_3}{g_3^2}~,
\eeqa
which predict that $M_1:M_2:M_3\approx 1:2:6$ at the electroweak (EW) scale. For gluino mass heavier than 2.2 TeV, the bino and winos cannot be very light with such a gaugino mass ratio. Besides, the requirement to interpret the 125 GeV Higgs require the stop masses to be of order 5 TeV, given the vanishing trilinear coupling $A_t$ at the messenger scale and not very high messenger scale. Consequently, lower bounds will be set for the slepton masses at the EW scale by the mass relations among the sfermion masses at the messenger scale. Similarly, stringent LHC constraints on the masses of colored squarks for the first two generations also set stringent constraints on the slepton masses at the EW scale. The DM in gauge mediation is always the light gravitino whose relic abundance relies on the reheating temperature. So, we need not worry about the DM constraints in our following discussions. Even though the stringent DM constraints are relaxed, it is not possible for MGM to explain the $\Delta a_\mu^{SUSY}$ anomaly in the $3\sigma$ range unless it is extended.

In the EOGM, after imposing a non-trivial R-symmetry, the most general superpotential involving $N$ numbers of messengers (in ${\bf 5}\oplus\bar{\bf 5}$ representation of $SU(5)$) can be written as~\cite{EOGM}
\beqa
W= {\cal{M}}_{ij}(X) \phi_i\tilde\phi_j =
 (\lambda_{ij}X+m_{ij})\phi_i\tilde\phi_j
\label{EGM:superpotential}
\eeqa
with ${\cal{M}}_{ij}(X)= \lambda_{ij} X+m_{ij}$ the messenger mass
matrix.  Because of the selection rules set by the R-symmetry, the messenger mass matrix satisfies a non-trivial identity~\cite{EOGM}
\beqa
\det{\cal{M}}= X^n G(m,\lambda),~~\qquad n = {1\over R(X)} \sum_{i=1}^N(2-R(\phi_i)-R(\tilde \phi_i)),
\eeqa
 with $n$ an integer satisfying $0\leq n\leq N$ and $G(m,\lambda)$ some function of the couplings. After integrating out the messengers, the soft SUSY breaking parameters can be obtained as
\beqa
 M_r = \f{\alpha_r}{4\pi}\Lambda_G,~~~~
\Lambda_G = F\f{\pa}{\pa X}\(\f{}{}\log\det{\cal{M}}\)=\f{n F}{X}~,
\label{EGM:gaugino}
\eeqa
for gauginos and
\beqa
 m_{\tl{f}}^2 = 2 \sum_{r=1}^3 C_{\tl{f}}^r\left({\alpha_r\over 4\pi}\right)^2 \Lambda_S^2,\qquad
 \Lambda_S^2 = \f{1}{2} |F|^2 {\partial^2\over\partial X\partial
X^*}\sum_{i=1}^{N} \left(\log |{\cal{M}}_i|^2\right)^2~,
 \eeqa
for sfermion masses, where ${\cal{M}}_i$ denote the eigenvalues of $\cal{M}$. Vanishing trilinear couplings are predicted at the messenger scale.

The $''$\emph{effective messenger number}$''$ is defined as~\cite{EOGM}
\beqa
N_{\rm eff}(X,m,\lambda)
\equiv {\Lambda_G^2\over\Lambda_S^2}~,
\eeqa
which is a continuous function of the couplings taking values between $0$ and $N$ inclusive.
As large contributions to $\Delta a_\mu$ requires light sleptons and light eletroweakinos, we need to seperate the masses of the colored sparticles from the uncolored ones to evade the stringent collider constraints. The general form of superpotential eq.(\ref{EGM:superpotential}) can be amended to respect only the SM gauge symmetry instead of SU(5), which is
\beqa
W &=(\lambda_{2ij} X+m_{2ij}) \ell_i \tilde{\ell}_j+ (
\lambda_{3ij} X+m_{3ij} ) q_i \tilde{q}_j~.
\label{EGM:N3N2}
\eeqa
Similarly to the $SU(5)$ case, the sfermion masses are given as
\beqa
m_{\tilde f}^2 = 2 \sum_{r=1}^3 C_{\tilde
f}\,^r\left({\alpha_r\over 4\pi}\right)^2 \Lambda_{S;r}^2~,~~~~~~ \left\{\bea{l}\Lambda_{S;r=2,3}^2= \Lambda_{G}^2 N_{\rm eff}(X,m_r,\lambda_r)^{-1}~, \\
     ~~~~~~\Lambda_{S1}^2={2\over5}\Lambda_{S3}^2+{3\over5}\Lambda_{S2}^2~,
\eea \right.
\eeqa
 and the gaugino masses still take the form eq.(\ref{EGM:gaugino}).

It is possible to choose proper forms of the superpotential~(\ref{EGM:N3N2}) so as that $N_{eff,3}\ll N_{eff,2}$. In this case, the colored sfermions, which get dominant contribution of the form $\al_3^2/N_{eff,3}$, can be much heavier than the uncolored ones that are dominated by $\al_2^2/N_{eff,2}$ or $\al_1^2/N_{eff,3}$.  Although spectrum of this type is welcome to solve the muon $g-2$ anomaly and accommodate the 125 GeV Higgs, vacuum stability bounds will set very stringent constraints on such a type of spectrum. Besides, as the gaugino mass ratios $M_1:M_2:M_3\approx 1:2:6$ still hold in this EOGM model at the EW scale, the lightest gaugino is always the lightest bino, which is bounded below to be heavier than 350 GeV for 2.2 TeV lower bound on gluino mass. 

In ordinary GMSB, the right-handed (RH) sleptons should be much lighter than the left-handed (LH) ones because of the hierarchy from gauge couplings $\al_2^2\gg \al_1^2$. In EOGM with $N_{eff,3}\ll N_{eff,2}$, we therefore will have $N^{-1}_{eff,1}\gg N^{-1}_{eff,2}$ because of the relation
 \beqa
N^{-1}_{eff,1}=\f{3}{5}N^{-1}_{eff,2}+\f{2}{5}N^{-1}_{eff,3}~.
\eeqa
Such a hierarchy can compensate the hierarchy from gauge couplings $\al_2^2\gg \al_1^2$. So, LH sleptons can possibly be lighter than the RH ones in this case, which may give large contributions to W boson mass.

As gauge mediation does not address the $\mu$-problem, the generation of $\mu$-term needs an alternative generation mechanism, see~\cite{Bae:2019dgg} for a review of the solutions to $\mu$ problem in gauge mediation. Without giving concrete models on the generation of $\mu$ terms, the value of $\mu$ can be determined by the EWSB conditions.
One of the EWSB condition can be written as
\beqa
 \f{M_Z^2}{2}&=&\f{m^2_{H_d}-m_{H_u}^2\tan\beta^2}{\tan^2\beta-1}-\mu^2+{\cal O}(\f{1}{\tan^4\beta})~,\nn\\
             &\approx& -m_{H_u}^2(m_{\tl{t}})-\mu^2.~~~~~{\rm for~large~\tan\beta}~,\nn\\
             &\approx& -\(m_{H_u}^2(M_{mess})-\f{3 y_t^2}{4\pi^2}m_{\tl{t}}^2\log\(\f{M_{mess}}{m_{\tl{t}}}\)\)-\mu^2~.
\label{EWSB}
 \eeqa
Unlike the argument in~\cite{EOGM}, no large cancelation will occur for $N_{eff,3}\ll N_{eff,2}$. Therefore, the value of $\mu$ at the EW scale should lie of order
\beqa
\mu\sim \f{\al_3}{4\pi^2}\f{\Lambda_G}{\sqrt{N_{eff,3}}}~,
\eeqa
which in general is larger than the value allowed by EWSB in ordinary MGM.
Large value of $\mu$ means that the $\tl{\mu}_L-\tl{\mu}_R-\tl{B}$ contribution to $\Delta a_\mu$ is enhanced via  smuon mixing off-diagonal element.

\subsection{Numerical results the SUSY contributions to $\Delta a_\mu$ and $\Delta m_W$ in ordinary EOGM }
We try to explain the recent muon $g-2$ measurements with the EOGM scenario when $N_{eff,3}\ll N_{eff,2}$. The following free parameters are adopted as inputs at the messenger scale for our numerical scan
\beqa
\Lambda_G,~ ~0\leq N_{eff,3}\leq 1,~~ 1\leq N_{eff,2}\leq 6, ~~M_{mess}~,10<\tan\beta<50.
\eeqa
 In our scan, we impose the following constraints in addition to the constraints already encoded in the packages
  \bit
 \item[(i)] The lightest CP-even Higgs boson should act as the SM-like Higgs boson with a mass of $125\pm 2$ GeV~\cite{ATLAS:higgs,CMS:higgs}.
  \item[(ii)] Direct searches for low mass and high mass resonances at LEP, Tevatron and LHC by using the package HiggsBounds-5.5.0~\cite{higgsbounds511} and HiggsSignals-2.3.0~\cite{higgssignal}.
  \item[(iii)] Constraints on gluino and squark masses from the LHC~\cite{LHCmass}:
  \beqa
  m_{\tl{g}}\gtrsim 2.2 ~{\rm TeV},~~~~m_{\tl{q}}\gtrsim 1.4 ~{\rm TeV},
  \eeqa
and the lower mass bounds of charginos and sleptons from the LEP~\cite{LEPmass}:
  \beqa
  m_{\tl{\tau}}\gtrsim 93.2 ~{\rm GeV},~~~~  m_{\tl{\chi}^\pm} \gtrsim 103.5 ~{\rm GeV}.
  \eeqa
  \item[(iv)] Constraints from $B \to X_s \gamma$, $B_s \to \mu^+ \mu^-$and $B^+ \to \tau^+ \nu_\tau$ etc \cite{BaBar-Bph, LHCb-BsMuMu, Btaunu}
      \begin{eqnarray}
        3.15\times10^{-4} <&Br(B_s\to X_s \gamma)&< 3.71\times10^{-4} \\
         1.7\times10^{-9} <&Br(B_s\to\mu^+\mu^-)&< 4.5\times10^{-9}  \\
        0.78\times10^{-4} <&Br(B^+\to\tau^+\nu_\tau)&< 1.44\times10^{-4}~.
      \end{eqnarray}
  \item[(v)]  Vacuum stability bounds on the soft SUSY breaking parameters.

   Although it is attractive to require the ordinary EWSB vacuum to be the global minimum of the theory,  it is also acceptable that a deeper minima exist but the lifetime of the ordinary false EWSB vacuum decaying into the charge-breaking true vacuum via quantum tunneling is larger than the age of the universe. The following set of vacuum stability constraints are imposed:
\bit
\item   the vacuum meta-stability condition fitting formula given by~\cite{Kitahara:2013lfa}
\beqa
|\mu \tan \beta_{\textrm{eff}}| &<& 56.9 \sqrt{m_{\tilde{L}} m_{\tilde{\tau }R}} + 57.1 \left(m_{\tilde{L}}+1.03 m_{\tilde{\tau }R} \right)   - 1.28 \times 10^4 \GeV \nn\\
&+&\frac{1.67 \times 10^6 \GeV ^2 }{m_{\tilde{L}}+m_{\tilde{\tau }R} }  - 6.41 \times 10^7 \GeV ^3 \left ( \frac{1}{m_{\tilde{L}}^2  } + \frac{0.983}{m_{\tilde{\tau }R}^2}  \right) ~,
\label{metabound}
\eeqa
which can be applied in the region where  $m_{\tilde{L}}$, $m_{\tilde{\tau }_R} \leq 2$ TeV, the error of this fit being  less than 1 \% in this region. 
\item  An alternative approximate semi-analytic bound on CCB vacuum is given by
\begin{equation}
	A_t^2+3\mu^2 < (m_{\tl{t}_R}^2 + m_{\tl{t}_L}^2)\cdot
	\begin{cases}
		3\quad   & \text{stable,}     \\
		7.5\quad & \text{long-lived.}
	\end{cases}\label{eq:sanabounds}
\end{equation}
\eit
\eit

The SUSY contributions to the W-boson masses can be calculated with the Peskin's $S,T,U$ oblique parameters~\cite{STU,STU1,STU2}, which can be obtained at one-loop level with the package SPheno-4.0.5~\cite{Spheno}.
Knowing the oblique parameters, one can obtain the corresponding corrections to various electroweak precision observables. The shift of W-boson mass by new SUSY contributions at one-loop level can be given in terms of the $S,T,U$~\cite{STU,W:STU} parameters
\beqa
\Delta m_W=\f{\al M_W}{2(c_W^2-s_W^2)}\(-\f{1}{2}S+c_W^2 T+\f{c_W^2-s_W^2}{4s_W^2} U\)~,
\eeqa
with
\beqa
\alpha S & = & 4s_w^2 c_w^2
               \left[ \AP0{ZZ}
                          -\frac{c_w^2-s_w^2}{s_w c_w}\AP0{Z\gamma}
                          -\AP0{\gamma\gamma}
               \right]\,,  \nonumber \\
\alpha T & = & \frac{\A0{WW}}{m_w^2} - \frac{\A0{ZZ}}{m_Z^2}\,, \\
\alpha U & = & 4s_w^2
               \left[ \AP0{WW} - c_w^2\AP0{ZZ}
                         - 2s_wc_w\AP0{Z\gamma} - s_w^2\AP0{\gamma\gamma}
               \right]\,, \nonumber
\eeqa
and $\al^{-1}(0)=137.035999084~,s_W^2=0.23126$.

\begin{figure}[htb]
\begin{center}
\includegraphics[width=2.9in]{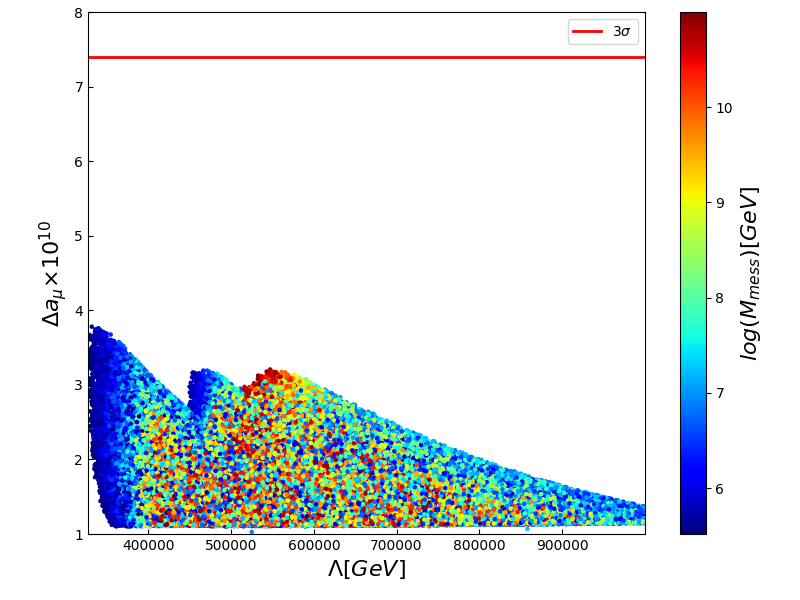}
\includegraphics[width=2.9in]{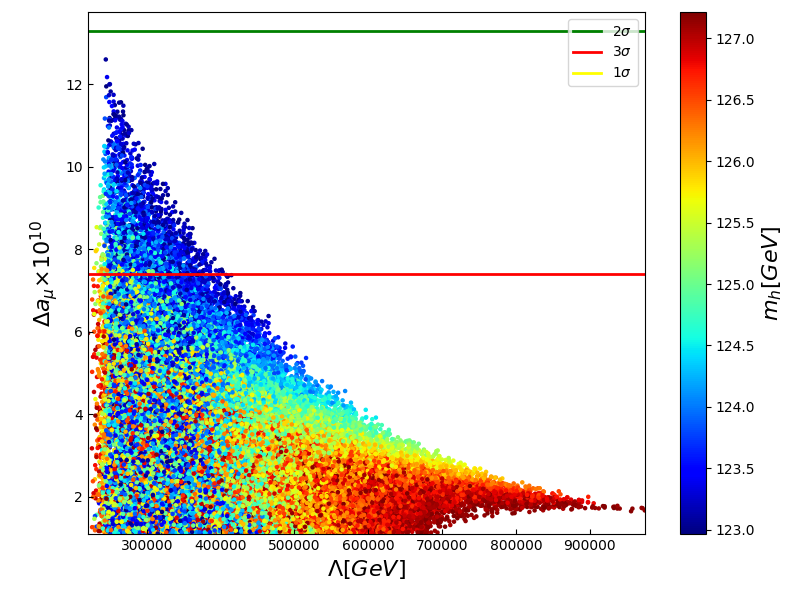}\\
\includegraphics[width=2.9in]{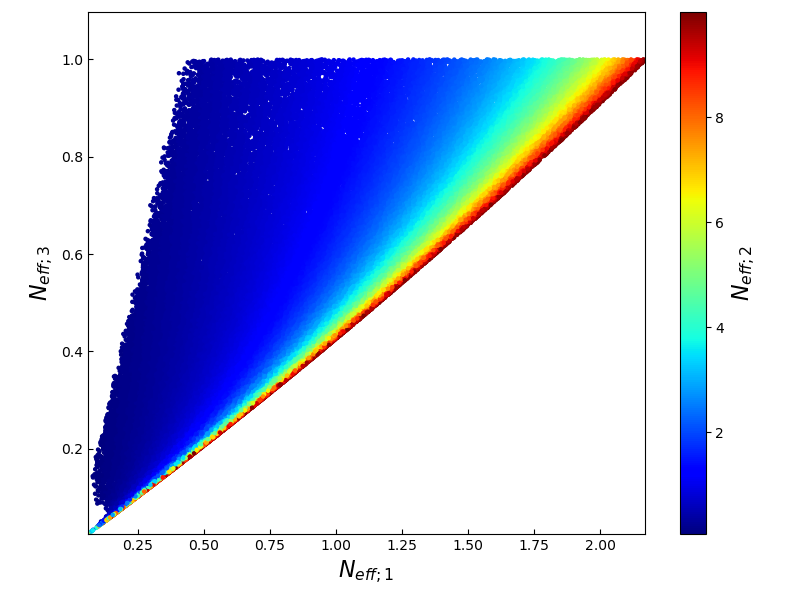}
\includegraphics[width=2.9in]{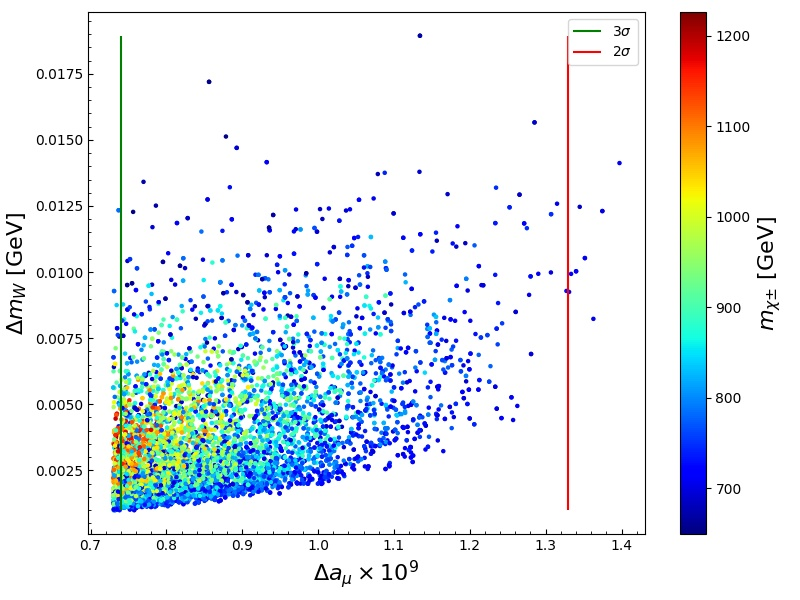}\\
\vspace{-.5cm}\end{center}
\caption{ Survived points that can satisfy the constraints (i-v) and give SUSY contributions to $\Delta a_\mu$ up to the $3\sigma$ range of $\Delta a_\mu^{combine}$. To compare the results, we show in the upper left panel the SUSY contributions to $\Delta a_\mu$ in ordinary GMSB. In the upper right panel, the SUSY contributions to $\Delta a_\mu$ in EOGM are shown. In the lower right panel, the SUSY contributions to $\Delta m_W$ in EOGM are shown with the corresponding ranges of $\Delta a_\mu$. }
\label{fig1}
\end{figure}

Our numerical results are shown in fig.\ref{fig1}. To compare the EOGM scenario with ordinary GMSB, we show in the upper left panel the SUSY contributions to $\Delta a_\mu$ in ordinary GMSB. It can be seen that ordinary GMSB can hardly lead to large $\Delta a_\mu$ upon $3\sigma$.
In EOGM, it can be seen from the upper-right panel that the muon $g-2$ anomaly can be explained up to $3\sigma$ range. Although such a result is still not satisfactory, it nevertheless improve the situation of ordinary GMSB, which can hardly reach the $3\sigma$ range. In our EOGM scenario, the gaugino mass ratios still set an lower bound on the lightest electroweakino mass, just as that in MGM. On the other hand, heavy colored sparticles required by collider exclusion bounds and 125 GeV Higgs masses still allow much lighter sleptons because the mass relations in MGM between squarks and sleptons are relaxed in our EOGM scenario with $N_{eff,3}\ll N_{eff,2}$.

We should note that the vacuum stability bound eq.(\ref{metabound}) can set very stringent constraints on the otherwise allowed parameter spaces because the value of $\mu$ in our EOGM scenario is large. In fact, without taking into account the vacuum stability bounds, the SUSY contributions to $\Delta a_\mu$ can reach the 1$\sigma$ range in this EOGM scenario.

The SUSY contribution to W-boson mass $\Delta m_W$ in this EOGM scenario is shown in the lower right panel. We anticipate that relatively light LH sleptons may possibly give sizeable contributions to $\Delta m_W$. However, it can be seen from the panel that the SUSY contributions to $m_W$ can at most reach $0.018$ GeV in the region the muon $g-2$ reached $3\sigma$. Such a sizable SUSY contribution to $m_W$ can not explain the new CDF II data.

\section{\label{sec-3}The solution to $g_\mu-2$ in adjoint messenger extended EOGM model}
The gaugino mass ratios $M_1:M_2:M_3\approx 1:2:6$ in EOGM set stringent constraints on the electroweakino masses, which cause some troubles in explaining the muon $g-2$ anomaly. So, it is interesting to extend the EOGM so as that such gaugino mass ratios are spoiled.
We can introduce $N$ families of $\Psi_a({\bf  5})\oplus \overline{\Psi}_a({\bf \overline{5}})$ representation messengers as well as $n_{\Sigma}$ pairs of $\Sigma_{3;b}({\bf 1,3,0})\oplus \Sigma_{8;b}({\bf 8,1,0})$ messengers at an common adjoint messenger scale $m_{adj}$\footnote{The octet $\Sigma_{8;b}$ and triplet $\Sigma_{3;b}$ messengers in general should take different scales to guarantee successful gauge coupling unification. As their scales lie at almost the same scale, a common scale of order $M_{Pl}^{2/3}m_{SUSY}^{1/3}$, which lies between $10^{12}$ GeV and $10^{14}$ GeV, can be used.}. As noted in~\cite{Mess:adjoint,HT:adjoint}, such adjoint messengers will not spoil the gauge coupling unification with the unification scale at the string scale $M_{str}\approx 5.27\tm 10^{17}$ GeV.
The superpotential involving both types of messengers are given as
\beqa
W\supseteq \(\la_{ab} X+m_{ab}\) \overline{\Psi}_a\Psi_b+\la X \(\Sigma_{3;c}\Sigma_{3;c}+\Sigma_{8;d}\Sigma_{8;d}\).
\eeqa
It is also possible for the adjoint messenger sector to have the EOGM type superpotential. However, as the adjoint messengers can give large contributions to the beta functions of the gauge couplings, it is better to introduce as small numbers of adjoint messengers as possible to avoid strong gauge couplings at the GUT scale. So we just adopt the ordinary MGM-type messenger sectors for adjoint messengers.
After integrating out $\Psi_a,\overline{\Psi}_a$ and $\Sigma_{3;b},\Sigma_{8;b}$, we can get the soft SUSY breaking gaugino masses at the messenger scale
\beqa
M_1&=&\f{\al_1}{4\pi}\f{F}{M}N_\phi\equiv \f{\al_1}{4\pi}\Lambda_G ~,\nn\\
M_2&=&\f{\al_2}{4\pi}\f{F}{M}\(N_\phi+2n_\Sigma\)\equiv \f{\al_2}{4\pi}\Lambda_G\(1+\f{2n_\Sigma}{N_\phi}\) ~,\nn\\
M_3&=&\f{\al_3}{4\pi}\f{F}{M}\(N_\phi+3n_\Sigma\)
\equiv \f{\al_3}{4\pi}\Lambda_G\(1+\f{3n_\Sigma}{N_\phi}\)~,
\eeqa
with $N_\phi$ an integer satisfying $0\leq N_\phi\leq N$ and $\Lambda_G\equiv N_\phi {F}/{M}$. Therefore, the gaugino ratios change approximately as $M_1:M_2:M_3\approx N_\phi:2\(N_\phi+2n_\Sigma\):6 \(N_\phi+3n_\Sigma\)$ at the EW scale. With the spoiled gaugino ratios, the bino can be even lighter than the previous EOGM case, which is welcome to explain the muon $g-2$ anomaly.

The soft scalars masses receive contributions from both types of messengers, which are given as
\beqa
m_{\tl{f}}^2&=&2\sum\limits_{r=1}^3 C_{\tl{f}}^r \(\f{\al_r}{4\pi}\)^2 \sum\limits_{a}\(\Lambda_{S,a}^2+D_r\Lambda_G^2\),
\eeqa
at the messenger scale, with $C_{\tl{f}}^r$ the corresponding Dynkin index and $D_r$ some coefficient for the contributions from adjoint messengers. Such expressions can be deduced from the discontinuity of the anomalous dimensions across the messenger thresholds  by the wavefunction renormalization method proposed in~\cite{Giudice:GMSB}.

Similar to ordinary EOGM, we can define
\beqa
N^{-1}_{eff,a}\equiv\f{\Lambda_{S,a}^2}{\Lambda_G^2}~.
\eeqa
The soft scalar masses are given as
\beqa
m_{H_u}^2&=&m_{H_d}^2=m_{L_L;i}^2=\f{3}{2}\(\f{\al_2}{4\pi}\)^2\[N^{-1}_{eff,2}+\f{2n_\Sigma}{N_\phi^2}\]\Lambda_G^2
+\f{3}{10}\(\f{\al_1}{4\pi}\)^2\[N^{-1}_{eff,1}\]\Lambda_G^2~,~\nn\\
m_{Q_L;i}^2&=&\f{8}{3}\(\f{\al_3}{4\pi}\)^2\[N^{-1}_{eff,3}+\f{3n_\Sigma}{N_\phi^2}\]\Lambda_G^2+
\f{3}{2}\(\f{\al_2}{4\pi}\)^2\[N^{-1}_{eff,2}+\f{2 n_\Sigma}{N_\phi^2}\]\Lambda_G^2
+\f{1}{30}\(\f{\al_2}{4\pi}\)^2\[N^{-1}_{eff,1}\]\Lambda_G^2~,~\nn\\
m_{U_L;i}^2&=&\f{8}{3}\(\f{\al_3}{4\pi}\)^2\[N^{-1}_{eff,3}+\f{3n_\Sigma}{N_\phi^2}\]\Lambda_G^2
+\f{8}{15}\(\f{\al_1}{4\pi}\)^2\[N^{-1}_{eff,1}\]\Lambda_G^2~,~\nn\\
m_{D_L;i}^2&=&\f{8}{3}\(\f{\al_3}{4\pi}\)^2\[N^{-1}_{eff,3}+\f{3n_\Sigma}{N_\phi^2}\]\Lambda_G^2
+\f{2}{15}\(\f{\al_1}{4\pi}\)^2\[N^{-1}_{eff,1}\]\Lambda_G^2~,~\nn\\
m_{E_L^c;i}^2&=&\f{6}{5}\(\f{\al_1}{4\pi}\)^2\[N^{-1}_{eff,1}\]\Lambda_G^2~,~\nn\\
\eeqa
with
\beqa
N^{-1}_{eff,1}=\f{3}{5}N^{-1}_{eff,2}+\f{2}{5}N^{-1}_{eff,3}~.\nn
\eeqa
We can see that ordinary hierarchical structures among sfermions in MGM persist because of the additional contributions from adjoint messengers. Hierarchy between $N_{eff,3}$ and $N_{eff,2}$ will no longer alter  the sfermion hierarchical structures. In this case, the soft SUSY breaking trilinear scalar couplings still vanish at the messenger scale.

 We would like to discuss the GUT constraints on the adjoint messenger extension EOGM scenarios.
The evolution of the gauge couplings, after neglecting the light MSSM and top threshold corrections as well as the heavy GUT scale threshold corrections are given by
\beqa
\al_i^{-1}(m_{GUT})&=&\al_i^{-1}(m_Z)-\f{b_i^{MSSM}}{2\pi}\ln\lrf{m_{GUT}}{m_{Z}}
-\f{1}{2\pi}\sum\limits_{n=1}^N\ln\lrf{m_{GUT}}{{\cal M}_{i;n}}-\f{n_\phi^i}{2\pi}\ln\lrf{m_{GUT}}{m_\phi},~\nn\\
&=&\al_i^{-1}(m_Z)-\f{b_i^{MSSM}}{2\pi}\ln\lrf{m_{GUT}}{m_{SUSY}}
-\f{N}{2\pi}\ln\lrf{m_{GUT}}{\overline{\cal M}_i}-\f{n_\phi^i}{2\pi}\ln\lrf{m_{GUT}}{m_{adj}}~,\nn\\
\eeqa
with $n_\phi^i=(0, 2n_\Sigma, 3n_\Sigma)$ and the MSSM gauge beta function $b_i^{MSSM}=(\f{33}{5},1,-3)$. The expression ${\cal M}_{i;n}$ is the $n$-th eigenvalue of the matrix ${\cal M}_i$ with $\overline{\cal M}_i\equiv(\det{\cal M}_i)^{1/N}$ and $\overline{\cal M}_1=(\overline{\cal M}_2)^{3/5}(\overline{\cal M}_3)^{2/5}$.

Because of the smaller running effects between the rather close scales $M_{adj}$ and $M_{string}$,
perturbativity requirement for the gauge couplings up to $M_{string}$ only leads to a loose bound on the numbers of the messenger states
\beqa
\f{N}{2\pi}\ln\lrf{m_{GUT}}{\overline{\cal M}}+\f{3 n_\Sigma}{2\pi}8.57\lesssim 29.6~,
\eeqa
with $\al(M_{string})\approx 1/20$ and ${\overline{\cal M}_2}\approx {\overline{\cal M}_3}\equiv \overline{\cal M}$. For $\overline{\cal M}\simeq m_{adj}=1.0\tm 10^{14}$ GeV, we have
\beqa
(3n_\Sigma+N)\lesssim 21.7~.
\eeqa
  However, the choices of $n_\Sigma$ are also constrained by the measure of unification
\beqa
B\equiv\f{\al_2^{-1}(m_Z)-\al_3^{-1}(m_Z)}{\al_1^{-1}(m_Z)-\al_2^{-1}(m_Z)}~.
\eeqa
Using the one-loop RGE for gauge couplings, we can obtain
\beqa
B=\f{\(b_2^{MSSM}-b_3^{MSSM}\)\ln\lrf{m_{GUT}}{m_{Z}}+N\ln\lrf{\overline{\cal M}_3}{\overline{\cal M}_2}-n_\Sigma\ln\lrf{m_{GUT}}{m_{adj}}}{\(b_1^{MSSM}-b_2^{MSSM}\)\ln\lrf{m_{GUT}}{m_{Z}}-\f{2}{5}N\ln\lrf{\overline{\cal M}_3}{\overline{\cal M}_2}-2n_\Sigma\ln\lrf{m_{GUT}}{m_{adj}}}~.
\eeqa
One-loop MSSM prediction for $B$ gives $B=5/7$, which agrees with experiment to approximately 5 percent accuracy. Requiring the possible deviation to lie within such a $5\%$ range will lead to the following constraints
\beqa
\left|N\ln\lrf{\overline{\cal M}_3}{\overline{\cal M}_2}+2.857n_\Sigma\right|\lesssim 5.65~,
\eeqa
after using $\ln(m_{GUT}/{m_{Z}})\approx 36.3$ and $\ln(m_{GUT}/{m_{adj}})\approx 8.57$ with the unification scale $m_{GUT}=M_{string}\approx 5.27\tm 10^{17}$ GeV. Taking into account the low scale and high scale threshold corrections to the evolution of gauge couplings in general can only relax the constraints a little bit. So, unless the relation ${\overline{\cal M}_3}\ll {\overline{\cal M}_2}$ holds, the value of $n_\Sigma$ should not be greater than 2.
\subsection{Numerical results the SUSY contributions to $\Delta a_\mu$ in extended EOGM with adjoint messengers}

With the spoiled gaugino mass relations,  we anticipate that the SUSY contributions to $\Delta a_\mu$ can be larger than that of the ordinary EOGM scenario.  We carry out numerical scan with the following free parameters as inputs
\beqa
\Lambda_G, ~0\leq N_{eff,3}\leq 1,~ 1\leq N_{eff,2}\leq 10, ~1\leq n_\Sigma \leq 2~, 2\leq N_\phi\leq 5~, 2\leq \tan\beta\leq 55,
\eeqa
at the messenger scale $m_{adj}=1.0\tm 10^{14}$ GeV. Constraints from (i) to (v) in the discussions for ordinary EOGM are also imposed for this extended EOGM with adjoint messengers. We should note that the value of $n_\Sigma$ can be larger than 2 only in case ${\overline{\cal M}_3}\ll {\overline{\cal M}_2}$. A larger $n_\Sigma$ will lead to a larger hierarchy between the gluino mass and electroweakino masses.

\begin{figure}[htb]
\begin{center}
\includegraphics[width=2.8in]{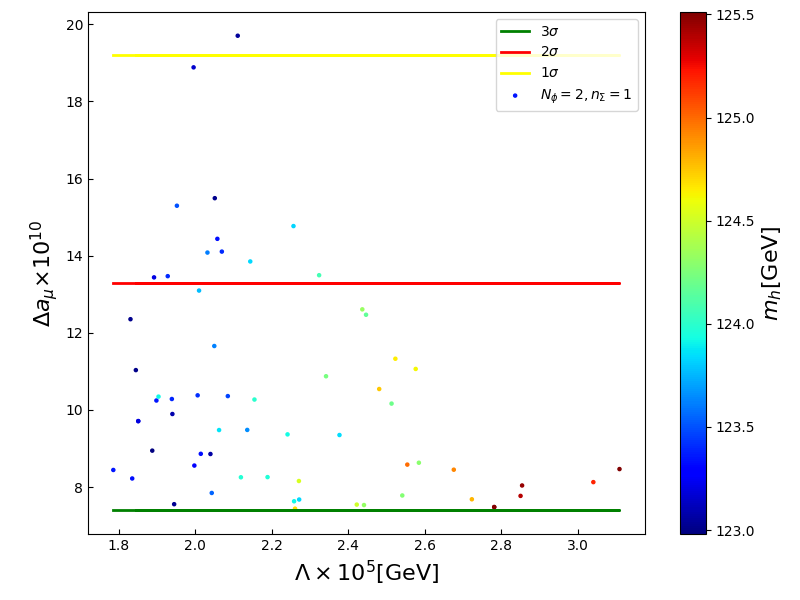}\includegraphics[width=2.8in]{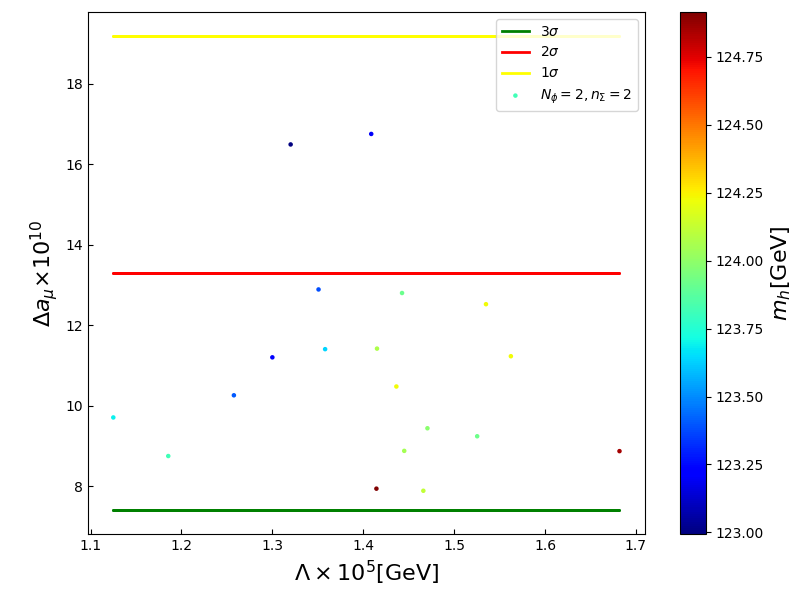}\\
\includegraphics[width=2.8in]{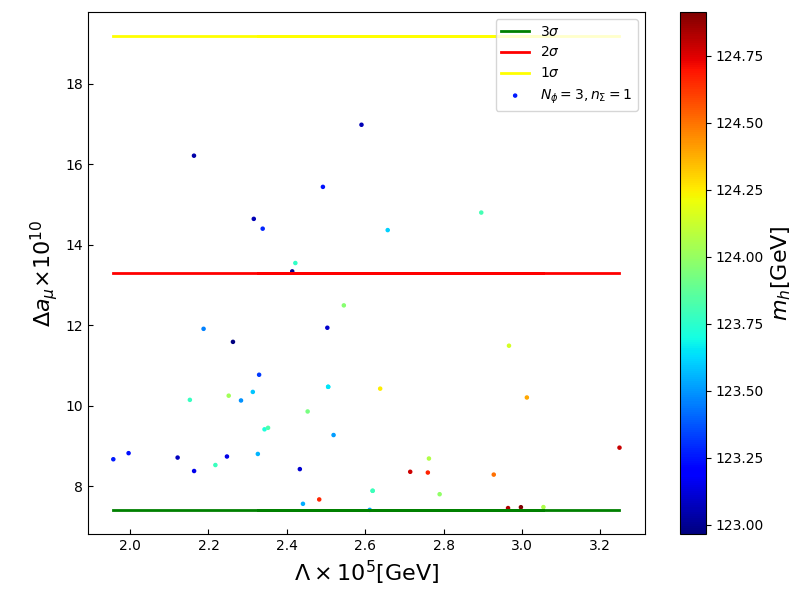}\includegraphics[width=2.8in]{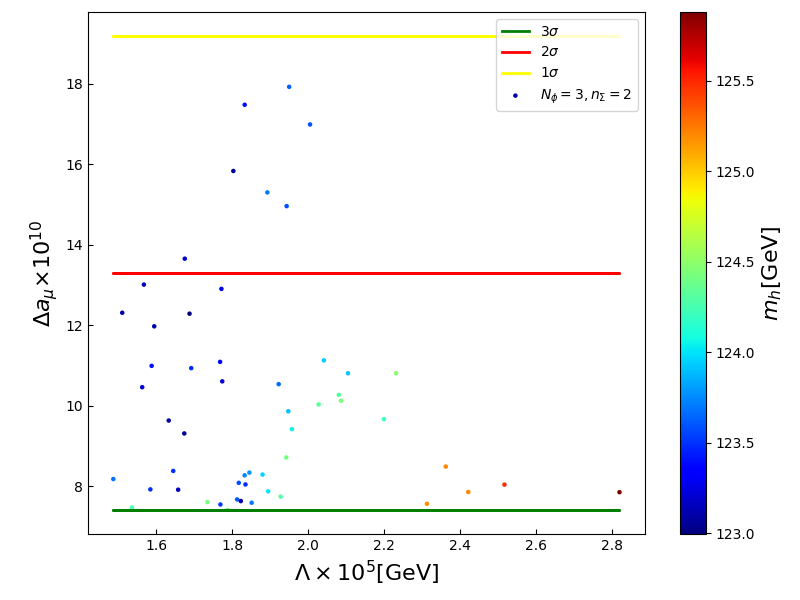}\\
\includegraphics[width=2.8in]{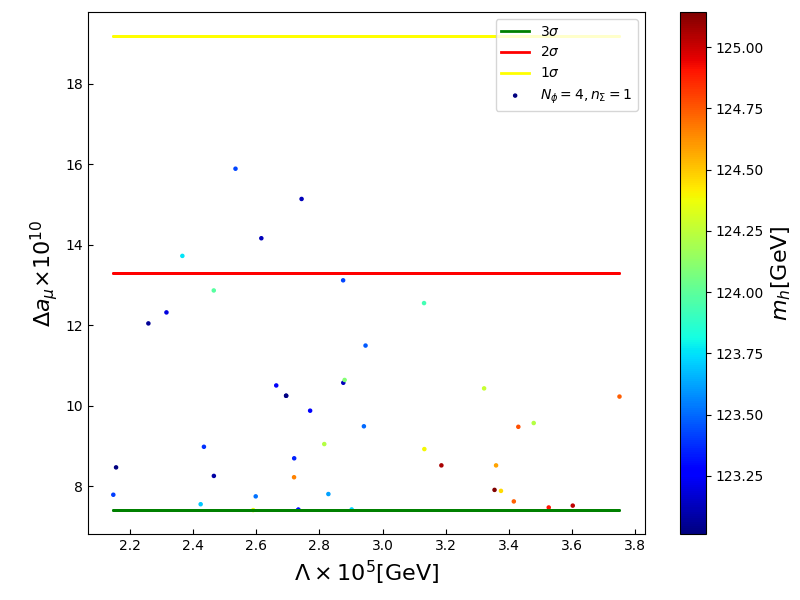}\includegraphics[width=2.8in]{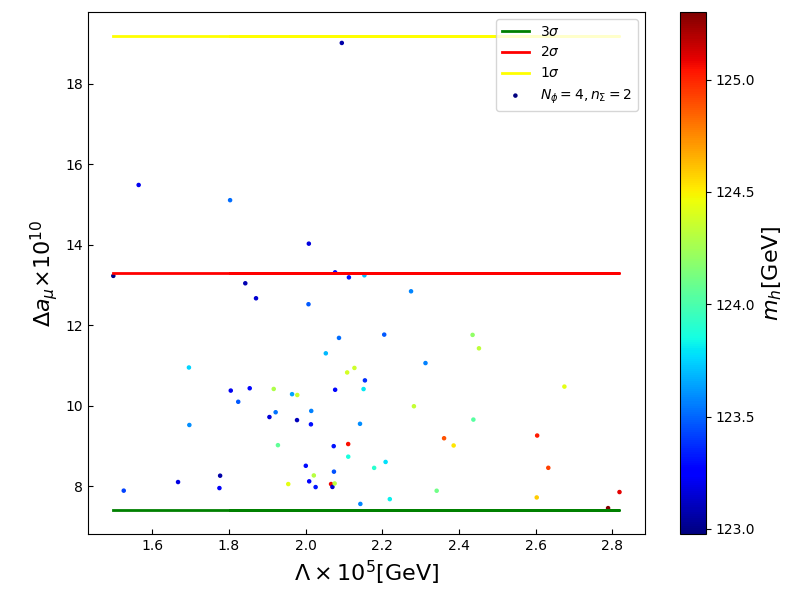}\\
\includegraphics[width=2.8in]{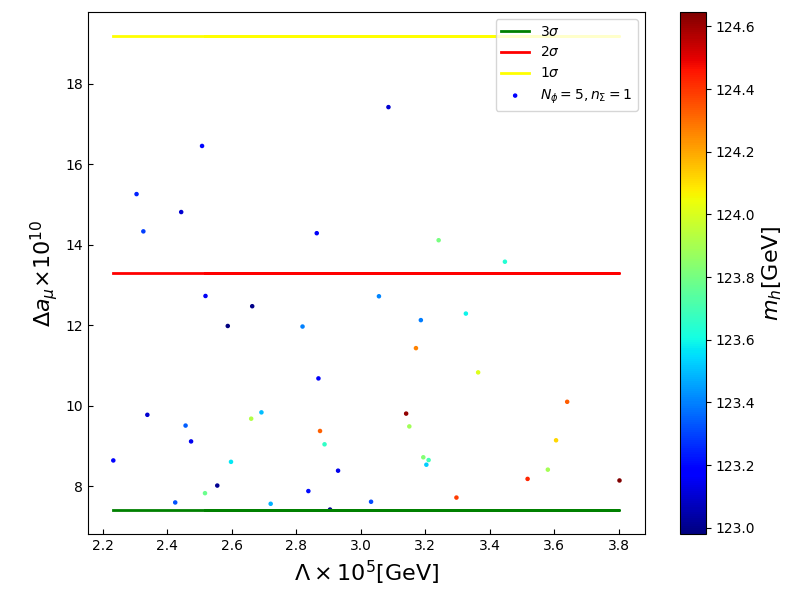}\includegraphics[width=2.8in]{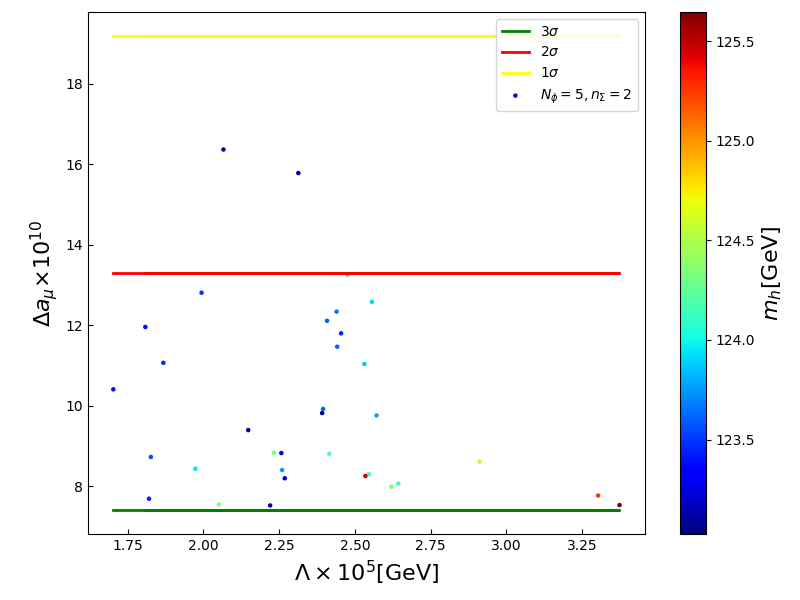}
\vspace{-.5cm}\end{center}
\caption{ Survived points that can satisfy the constraints (i) to (v) and explain the muon $g-2$ data upon $3 \sigma$ range. Their SUSY contributions to $\Delta a_\mu$ are shown in the vertical axis. The left panels (from top to bottom) correspond to $n_\Sigma=1$ and $N_\phi=2,3,4,5$, respectively. The right panels (from top to bottom) correspond to $n_\Sigma=2$ and $N_\phi=2,3,4,5$, respectively. We denote $\Lambda\equiv \Lambda_G$.}
\label{fig2}
\end{figure}

 We have the following discussions on our numerical results
\bit
\item The SUSY contributions to $\Delta a_\mu$ in various cases are shown in fig.\ref{fig2}.
It is obvious that such extended EOGM scenarios with adjoint messengers can explain the muon $g-2$ anomaly. All panels in fig.\ref{fig2} can reach the $2\sigma$ range of $\Delta a_\mu$. From previous discussions on gaugino mass ratios, it is clear that larger values of $n_\Sigma$ can relax further the stringent constraints from the 2.2 TeV lower bound of the gluino mass by LHC. Therefore, the wino and bino masses can take even smaller values, which is welcome to explain the muon $g-2$ anomaly.

The 125 GeV Higgs mass can also be accommodated in this scenario. Larger value of the predicted Higgs mass $m_h$, larger stop masses are in general required and consequently lead to larger slepton masses. So, larger $\Delta a_\mu$ in general prefers smaller $m_{h}$\footnote{Large value of $|A_t|$ at the EW scale can be generated from the messenger input $A_t=0$ with high messenger scale $M_{mess}\sim 1.0\tm 10^{14}$ GeV. With large $|A_t|$, the requirements of heavy stop masses by the 125 GeV Higgs mass can be relaxed. Consequently, the inverse dependence of $\Delta a_\mu$ on $m_{h}$ is not strict.}. Similarly, the mass parameter $\Lambda_G$ set the scale of the whole soft SUSY breaking spectrum. So, smaller value of $\Lambda_G$ leads to lighter sleptons and gauginos, which can predict larger $\Delta a_\mu$. In all cases, the gluino mass is constrained to lie below 6.9 TeV for those survived points that can explain the muon $g-2$ anomaly upon $3\sigma$ range.

The interaction strength for Goldstino component of gravitino is suppressed by $1/F$ instead of $1/M_{pl}$. Therefore, the decay of the next-to-lightest superpartner~(NLSP) to the LSP gravitino is not prompt unless $F$ is low. Superpartners of order TeV require that $F/M_{mess}\sim 100$ TeV, which can be used to determine the value of $F$ with $F\sim 1.0\tm 10^{19}$ GeV$^2$ for $M_{mess}=1.0\tm 10^{14}$ GeV.

 As discussed in~\cite{Draper:2011aa}, the decay length of NLSP is given by
\beqa
L=c\tau_{NLSP}\sim\f{16\pi^2 F^2}{m_{NLSP}^5}\approx 3\tm 10^{14} \(\f{100~{\rm GeV}}{m_{NLSP}}\)^5~m.
\eeqa
Thus, a $10^7$ GeV messenger scale corresponds to collider-size decay lengths, and higher messenger scales (for example, $m_{adj}=1.0\tm 10^{14}$ GeV in our cases) correspond to collider-stable NLSPs.
 With gluino mass of a few TeV, the NLSP lifetimes become ${\cal O}(10)$ seconds or longer. Such a long lifetime NLSP can play a significant role in post-BBN cosmology.

\begin{figure}[htb]
\begin{center}
\includegraphics[width=2.8in]{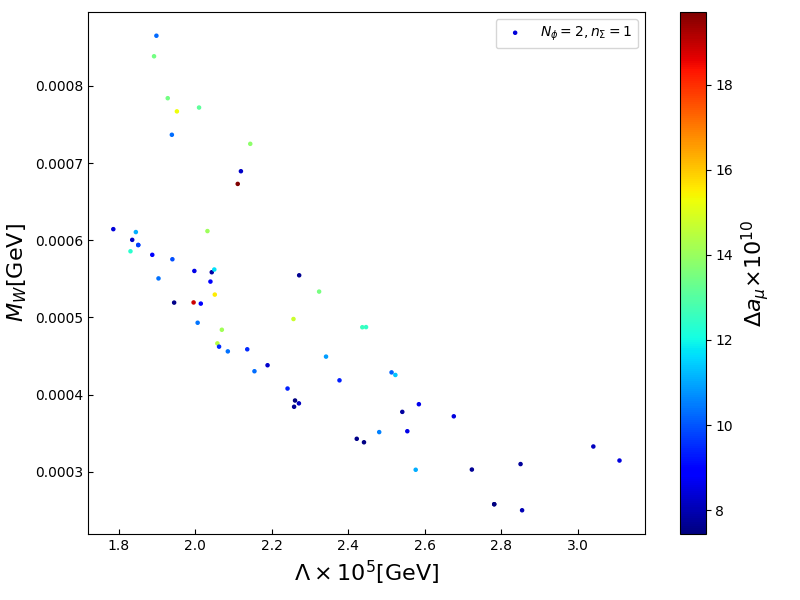}\includegraphics[width=2.8in]{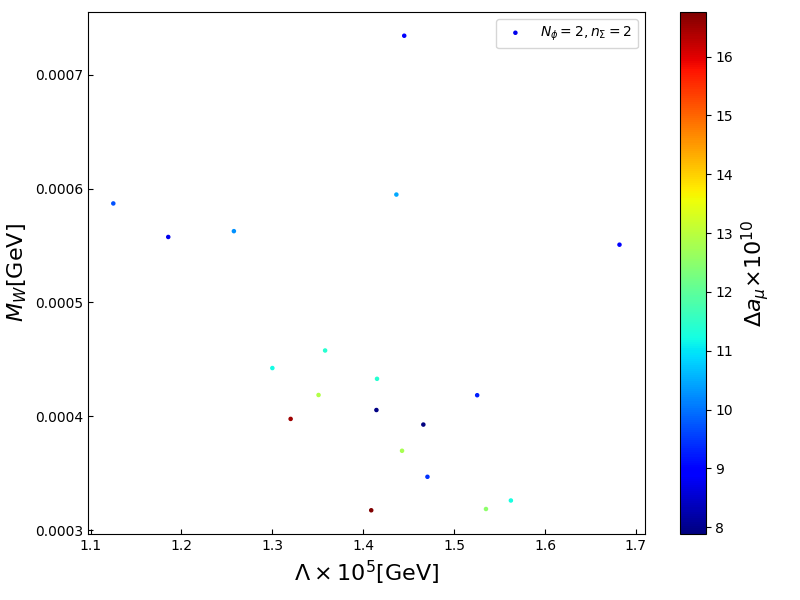}\\
\includegraphics[width=2.8in]{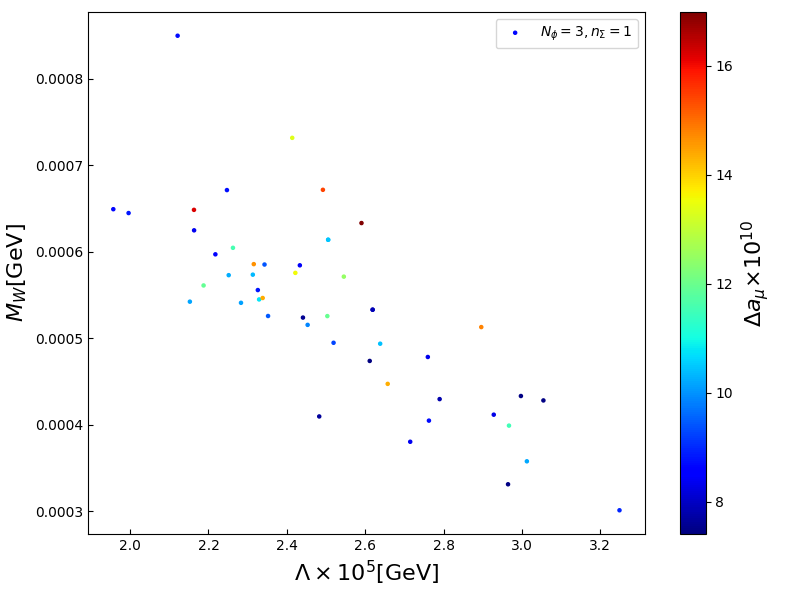}\includegraphics[width=2.8in]{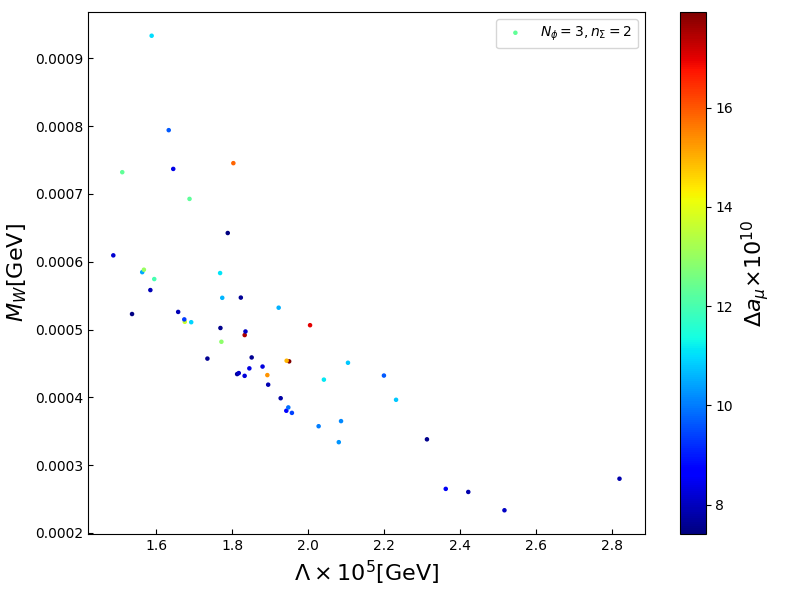}\\
\includegraphics[width=2.8in]{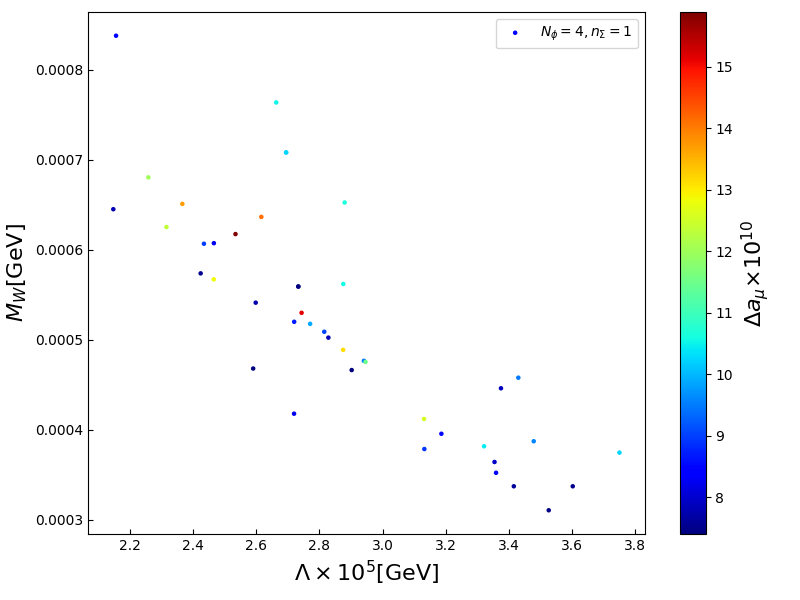}\includegraphics[width=2.8in]{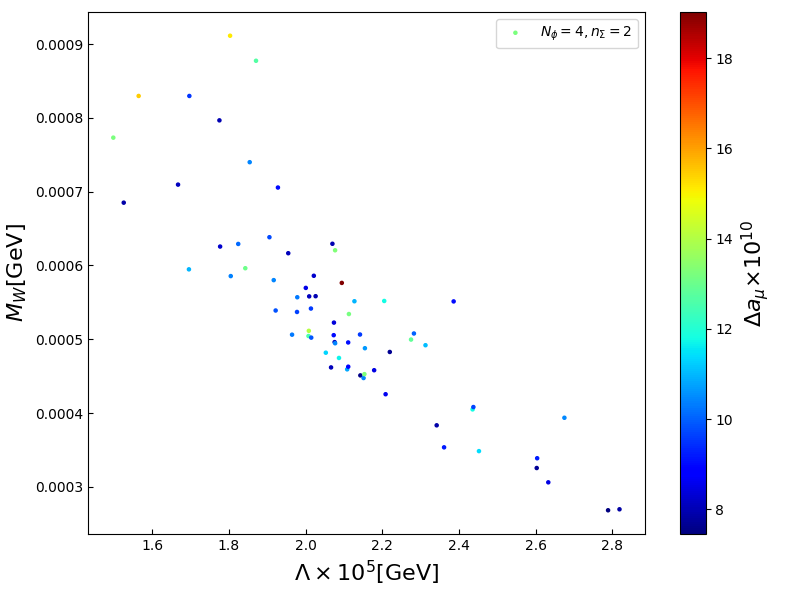}\\
\includegraphics[width=2.8in]{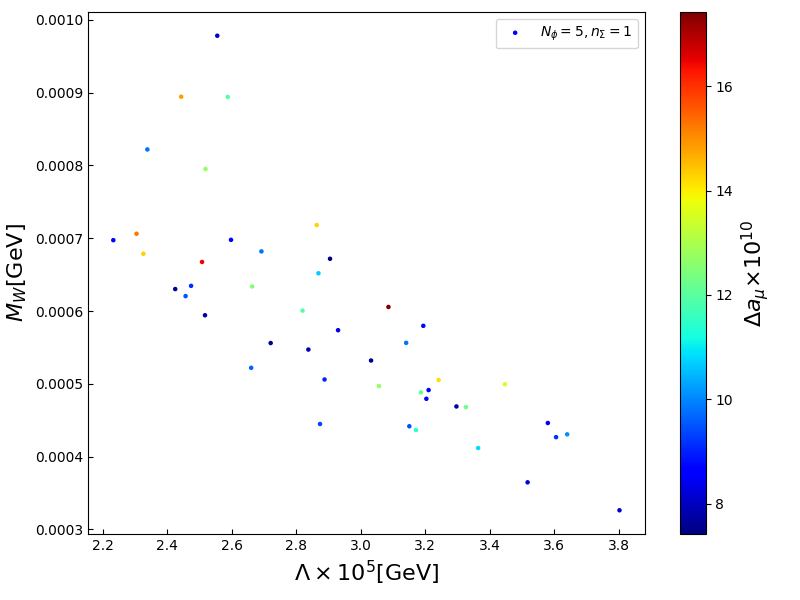}\includegraphics[width=2.8in]{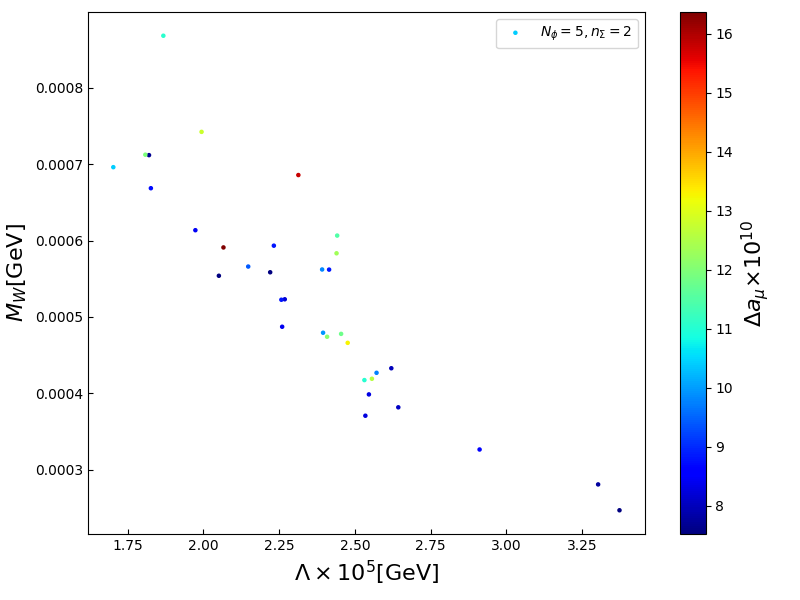}
\vspace{-.5cm}\end{center}
\caption{  Survived points shown in fig.2 and their SUSY contributions to $\Delta m_W$. The corresponding values of $\Delta a_\mu$ are also shown with different colors. The left panels (from top to bottom) correspond to $n_\Sigma=1$ and $N_\phi=2,3,4,5$, respectively. The right panels (from top to bottom) correspond to $n_\Sigma=2$ and $N_\phi=2,3,4,5$, respectively. }
\label{fig3}
\end{figure}

\item The SUSY contributions to W-boson mass are shown in fig.\ref{fig3} for each cases. It can be seen from the panels that the SUSY contributions to $\Delta m_W$ can at most reach 0.001 GeV, which is still too small to account for the recent CDF II data. It is also obvious that the SUSY contribution $\Delta m_W$ is inversely proportional to $\Lambda_G$ as $\Lambda_G$ set the scale of the soft SUSY breaking parameters. Lighter SUSY particles can in general lead to larger SUSY contributions to $\Delta m_W$. Lighter sleptons in these scenarios are always RH like, while wino and LH like sleptons are relatively heavy, leading to small contributions to $\Delta m_W$.
    
\begin{figure}[htb]
\begin{center}
\includegraphics[width=1.9in]{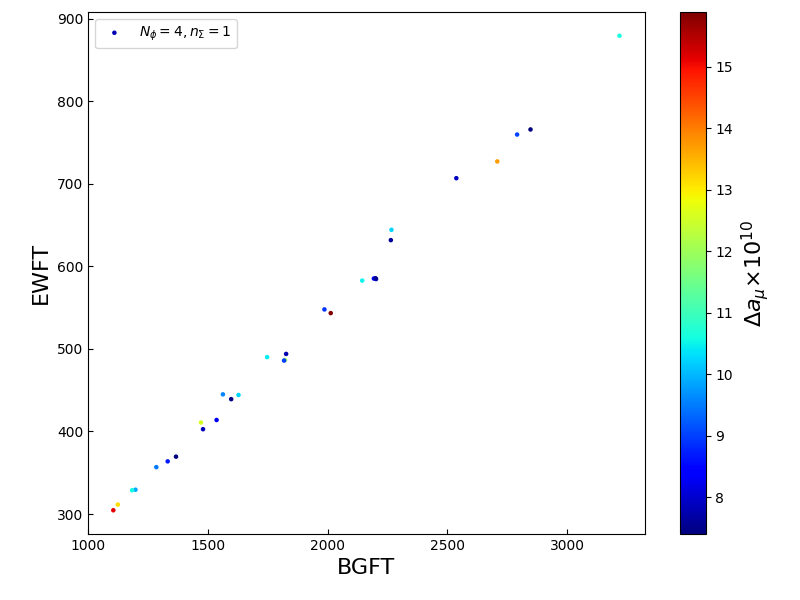}
\includegraphics[width=1.9in]{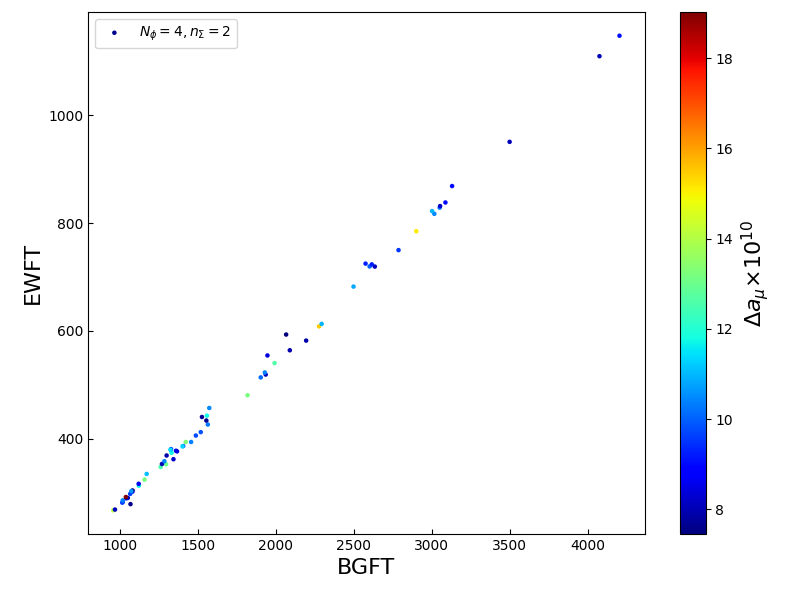}
\includegraphics[width=1.9in]{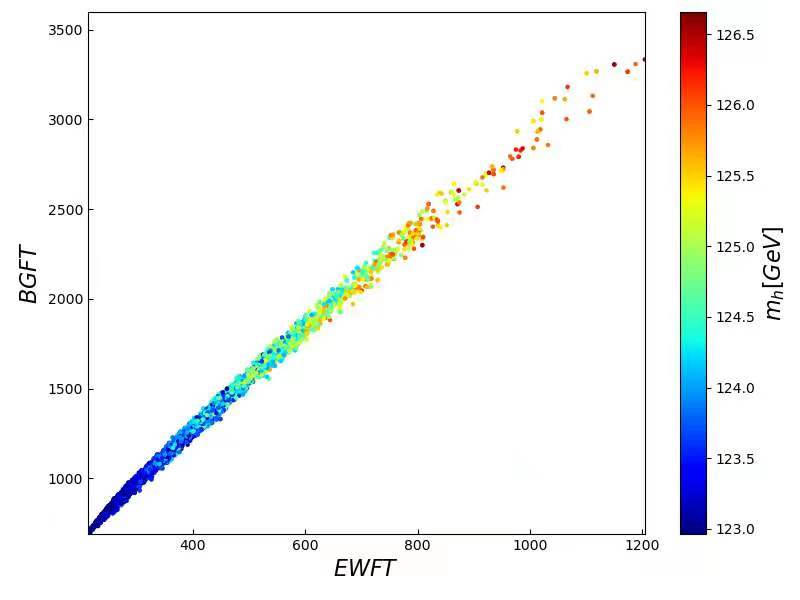}
\vspace{-.5cm}\end{center}
\caption{ Sample survived points that can satisfy the constraints (i) to (v) with their EWFT and BGFT measures for adjoint messenger extension of EOGM with $N_\phi=4,n_\Sigma=1$ (left panel), $N_\phi=4,n_\Sigma=2$ (middle panel) and EOGM (right panel), respectively. All the points shown in the adjoint messenger extension of EOGM scenarios can give  $\Delta a_\mu$ values upon $3\sigma$. No $3\sigma$ constraint are imposed to ordinary EOGM case. }
\label{fig4}
\end{figure}

\item We show the fine-tuning (FT) of the survived points in fig.\ref{fig4}. We use two types of FT measurements, the Barbieri-Giudice fine-tuning~(BGFT) measure $\Delta_{BG}$ and the electroweak fine-tuning~(EWFT) measure $\Delta_{EW}$.

     The total BGFT measure is defined as
          \beqa
          \Delta_{BG}= \max\limits_{i}\left|\f{\pa \ln M_Z^2}{\pa \ln a_i}\right|~,
          \eeqa
    with $\{i\}$ the set of parameters defined at the input scale. It was noted in~\cite{Baer:2013gva} that the BGFT will in general overestimate the FT involved. So, other FT measurements, such as the EWFT measure $\Delta_{EW}$, are also widely used in various studies. The EWFT measure $\Delta_{EW}$ is defined as~\cite{radiative:natural}
   \beqa
   \Delta_{EW}\equiv \max\limits_{i}(C_i)/\(\f{m_Z^2}{2}\),
   \eeqa
with
\beqa
C_{H_u}&=&\left|-\f{m^2_{H_u}\tan^2\beta}{\tan^2\beta-1}\right|~,~
C_{H_d}=\left|\f{m^2_{H_d}}{\tan^2\beta-1}\right|~,~
C_{\mu}=\left|-\mu_{eff}^2\right|~,~\nn\\
C_{\Sigma_{u}^u(\tl{t}_{1,2})}&=&
\f{\tan^2\beta}{\tan^2\beta-1}\left|\f{3}{16\pi^2}F(m_{\tl{t}_{1,2}}^2)\left[y_t^2-g_Z^2\pm \f{y_t^2 A_t^2-8g_Z^2(\f{1}{4}-\f{2}{3}x_w)\Delta_t}{m_{\tl{t}_{2}}^2-m_{\tl{t}_{1}}^2}\right]\right|~,\nn\\
C_{\Sigma_{d}^d(\tl{t}_{1,2})}&=&\f{1}{\tan^2\beta-1}\left|\f{3}{16\pi^2}F(m_{\tl{t}_{1,2}}^2)\left[g_Z^2\pm \f{y_t^2 \mu_{eff}^2+8g_Z^2(\f{1}{4}-\f{2}{3}x_w)\Delta_t}{m_{\tl{t}_{2}}^2-m_{\tl{t}_{1}}^2}\right]\right|~,
\eeqa
where $x_w=\sin^2\theta_W$ and
\beqa
\Delta_t&=&\f{(m_{\tl{t}_{L}}^2-m_{\tl{t}_{R}}^2)}{2}+M_Z^2\cos2\beta(\f{1}{4}-\f{2}{3}x_w)~,\nn\\
F(m^2)&=& m^2\(\log\f{m^2}{m_{\tl{t}_{1}}m_{\tl{t}_{2}}}-1\)~.
\eeqa
  In our numerical study, we give both the EWFT and BGFT measures for those survived points.

  The RGE evolution of $A_t$ from UV messenger scale to IR EW scale in general tends to a more and more negative value~\cite{Draper:2011aa} for positive gluino mass $M_3$. Given that the $A_t$ vanishes at the messenger scale, the value of $|A_t|$ in this scenario is in general larger than that of ordinary GMSB because of the much larger messenger scale $M_{mess}\simeq 1.0\tm 10^{14}$ GeV (hence a longer RGE evolution time). Large $|A_t|$ is welcome to accommodate the 125 GeV Higgs with TeV scale stop masses. Besides, large $A_t$ can also possibly lead to much smaller EWFT~\cite{radiative:natural} even for heavy stop masses. On the other hand, large value of $|A_t|$ will be constrained by the vacuum stability bound from fast decaying of meta-stable EWSB vacuum into color/charge breaking true vacuum.

In fig.\ref{fig4}, we show the BGFT and EWFT for those points that can survived all the constraints from (i) to (v) (including of course the vacuum stability bounds (v)).
  The FT for the adjoint messenger extension of EOGM cases with $N_\phi=4,n_\Sigma=1,2$ are shown in the left and middle panels, respectively. As anticipated, the EWFT can be as low as $200$ for stop masses of order TeV. All the points can explain the muon $g-2$ anomaly upon $3\sigma$. To compare with the FT of the ordinary EOGM case, we also show its EWFT and BGFT in the right panel. It can be seen that, the adjoint messenger extension of EOGM can indeed reduce the EWFT involved in comparison with ordinary EOGM because of its large messenger scale.

\eit
\subsection{The SUSY contributions to W-boson mass}
 From discussions in the previous EOGM sections, we can see that the SUSY contributions to W-boson mass can not account for the new CDF II data in EOGM and its adjoint messenger extension scenarios. 
So we would like to carry out more general discussions on possible SUSY contributions to W-boson mass in GMSB type scenarios. The chargino-neutralino fermion loop contribution to $\Delta m_W$ will have an opposite sign with respect to the LH slepton scalar loop contributions\footnote{The muon $g-2$ anomaly prefers light sleptons while the LHC bounds prefer heavy colored sparticles. So, squarks in general should be heavy, which are also naturally realized in GMSB type models.}. So, it is interesting to survey whether contributions of each type can give sizeable contributions, for example, light wino with heavy sleptons or light sleptons with heavy wino. In GMSB type scenarios, the DM constraints need no longer be imposed because the light gravitino DM sector is typically independent from the MSSM sector. So, some of the constraints for SUGRA/CMSSM type scenarios can be relaxed. Besides, the 125 GeV Higgs mass can be accommodated with ${\cal O}(10)$ TeV stop in GMSB for small $|A_t|$. Therefore, constraints from 125 GeV Higgs mass need no longer be imposed after assuming an independent heavy squark sector.

 We know that the gaugino mass ratios $M_1:M_2:M_3\approx 1:2:6$ are rather robust in gauge mediation scenarios. To obtain light wino of order 100 GeV with heavier than 2.2 TeV gluino mass, we need to increase the mass hierarchy between $M_3$ and $M_2$. However, in adjoint messenger extension of EOGM scenarios, larger $n_\Sigma$ can relax the gaugino ratios to at most  $M_3:M_2\approx 9:2$, which is still not enough to give very light wino. So we have to add additional mechanism to increase the gaugino hierarchy between $M_3$ and $M_2$ so as that light wino can be allowed (~for example, with product group proposed in~\cite{yanagida}). The relations (hierarchies) among the sfermion masses can also be spoiled with EOGM type messenger sector. So we can just discuss the most general SUSY spectrum without specifying concretely its UV (GMSB model building) origin.

\begin{figure}[htb]
\begin{center}
\includegraphics[width=2.8in]{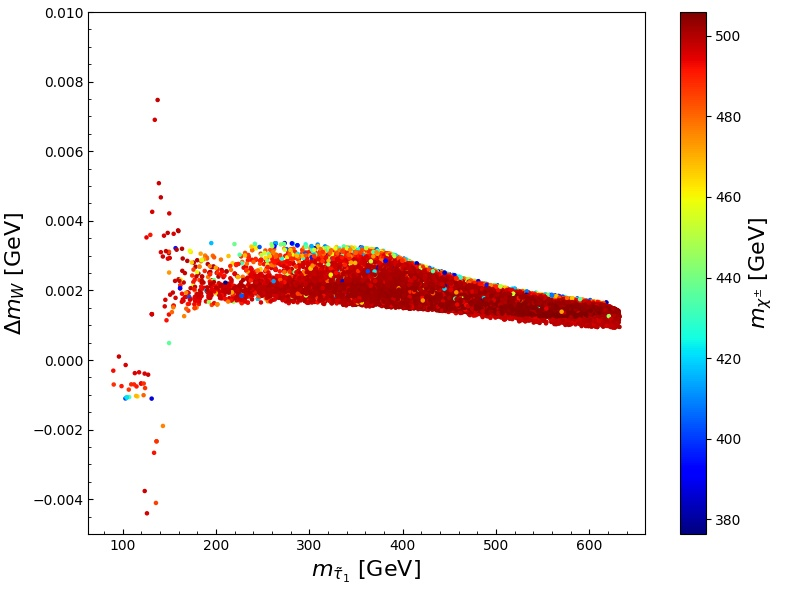}
\includegraphics[width=2.8in]{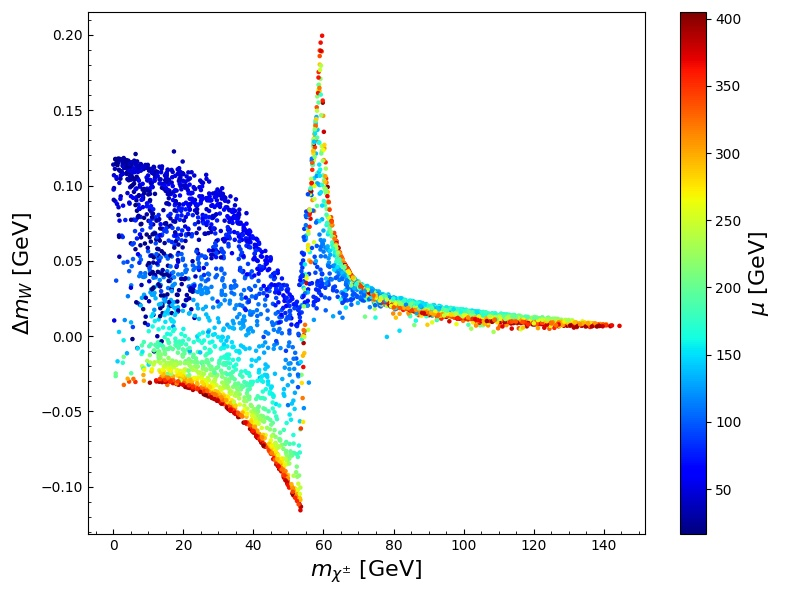}
\includegraphics[width=2.8in]{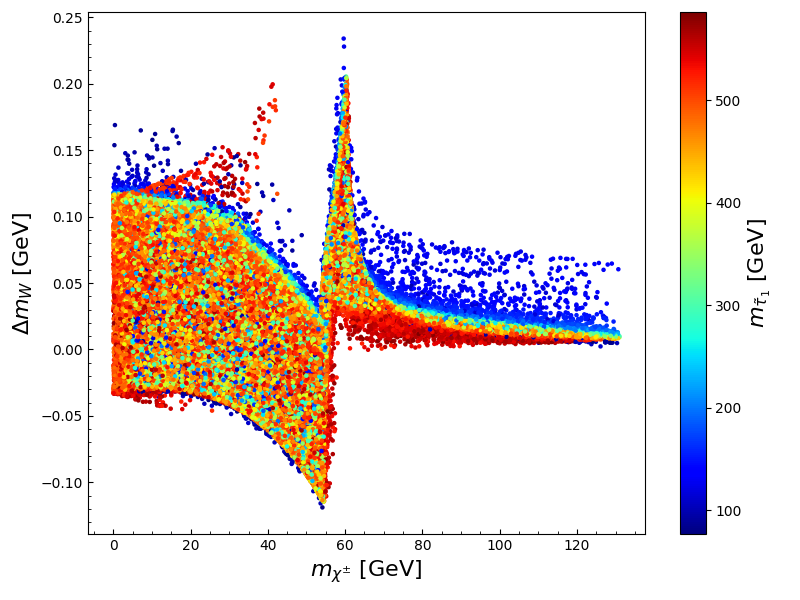}
\vspace{-.5cm}\end{center}
\caption{ The SUSY contributions to $\Delta m_W$ with  (I) light sleptons (left panel); (II) light wino (middle panel); (iii)both light wino and light sleptons (right panel), respectively.}
\label{fig5}
\end{figure}

 The SUSY contribution to $\Delta m_W$ for cases, with (I) light sleptons; (II) light wino; (III) both light wino and light sleptons, are shown in the panels of fig.\ref{fig5}, respectively. It can be seen that, in the light slepton case with heavy wino, the SUSY contribution to $\Delta m_W$ can reach at most to 0.008 GeV, which is too small to account for the CDF II data. In the light wino case with heavy sleptons, the SUSY contribution to $\Delta m_W$ can reach at most to 0.0165 GeV for chargino mass heavier than $103$ GeV, which still can not explain the CDF II data. In the case with both light wino and light sleptons, the SUSY contribution to $\Delta m_W$ can possibly give large $\Delta m_W$, which can reach up to $0.06$ GeV and possibly explain the new CDF II data.
 We check for those sample points in the large $\Delta m_W$ region that, after MC simulation, most points can still survive because of the very small mass difference between charginos and stau.

 So, scenarios with light sleptons only or light wino only can not explain the CDF II data on W-boson mass. When both sleptons and wino are light, GMSB type scenarios can marginally explain the new W-boson mass data reported by CDF II.

\section{\label{sec-4} Conclusions}
The SUSY contributions $\Delta a_\mu$ to muon $g-2$ anomaly can not even reach $3\sigma$ in ordinary GMSB scenarios because of the strong correlations between the colored sparticle masses and the EW sparticle masses. An interesting extension to GMSB is the EOGM, which can relax the correlations between squarks and sleptons with non-universal choices for $N_{eff,3}$ and $N_{eff,2}$. We find that EOGM scenarios with $N_{eff,3}\ll N_{eff,2}$ can explain the muon $g-2$ anomaly within $3\sigma$ range, however can not explain the new W-boson mass by CDF II. We also propose to extend EOGM with additional adjoint $\Sigma_8$ and $\Sigma_3$ messengers at a high scale of order $1.0\times 10^{14}$ GeV, which can shift the gauge coupling unification
scale to the string scale. Such EOGM extension scenarios with adjoint messengers could spoil the unwanted gaugino mass ratios and give large SUSY contributions to $\Delta a_\mu$ for $N_{eff,3}\ll N_{eff,2}$, which can explain the muon $g-2$ anomaly up to $1\sigma$. Besides, because of the large messenger scale of order $1.0\times 10^{14}$ GeV, such scenarios will in general lead to large $|A_t|$ at the EW scale, which can accommodate the 125 GeV Higgs easily and possibly lead to smaller EWFT as well as BGFT. The CDF II reported W boson mass can still not be explained in such EOGM scenarios with adjoint messengers extension. 

  We discuss the possibility to explain the new CDF II W-boson mass in general GMSB-type framework. Ordinary GMSB and EOGM extension models can not give large $\Delta m_W$ because of the strong correlations of the soft SUSY breaking spectrum. Relaxing some of the constraints, especially the gaugino ratios, the SUSY contributions can marginally account for the new W-boson mass data in the region with sleptons and wino both being light.
 
 AMSB type framework, which always predict light wino (being the lightest gaugino) and light sleptons, can be welcome to explain both the new CDF II data on W boson mass and the muon $g-2$ anomaly.

\begin{acknowledgments}
This work was supported by the National Natural Science Foundation of China (NNSFC) under grant Nos. 12075213, by the Key Research Project of Henan Education Department for colleges and universities under grant number 21A140025.
\end{acknowledgments}

\end{document}